\documentstyle[preprint,aps,prl]{revtex}
\begin{document}
\title{Higher Dimensional Schwinger-like Anomalous Effective Action 
\footnote{Accepted for publication on Phys.Rev.{\bf D}}}

\author{A.Smailagic \footnote{E-mail address: a.smailagic@etfos.hr} }
\address{Department of Physics, 
Faculty of Electrical  Engineering \\
University of Osijek, Croatia }
\author{E. Spallucci\footnote{E-mail address:spallucci@trieste.infn.it}}
\address{Dipartimento di Fisica Teorica\break
Universit\`a di Trieste,\\
INFN, Sezione di Trieste}
\maketitle

\begin{abstract}
 We construct explicit form of the anomalous effective action, in
{\it arbitrary even dimension }, for Abelian
vector and axial gauge fields coupled to Dirac  fermions. 
It turns out to be a surprisingly simple extension of $2$D Schwinger model
effective action.
\end{abstract}
\newpage

Problem of anomalies in field theory is an outstanding and old one. Since the
early days of their discovery \cite{adl},\cite{jackiw}, anomalies  
have been approached and investigated in many ways \footnote{Due to
 proliferation of literature on this subject, we are not able to quote all 
 papers containing important contributions. We apologize to the authors of 
 unquoted papers.}.  Only in two spacetime ($2$D) dimensions the  
corresponding anomalous effective action was calculated thanks to a particularly
simple, anomalous Feynman diagram. In this way famous Schwinger model
 and Liouville gravity anomalous actions were found \footnote{ 
 We have given a reference of one of a
 comprehensive reviews on the two-dimensional models.}
 \cite{berger}. Extending the calculation of the effective action from
  $2$D to $4$D one is faced  with considerable technical difficulties for the 
  explicit calculation of the { \it finite,} i.e. {\it non-local } part of the 
  anomalous triangle diagram. On the other hand, it is the non-local part of 
  the effective action\cite{adl},\cite{ros},
 which is unambiguous and  is the source of the anomalous behavior 
\cite{duff},\cite{gris}. Knowledge of the effective action, apart from giving 
a complete information on the theory, 
   has proven crucial in the recent study of the two-dimensional
 gauge and  gravitational anomalous theories in the light-cone gauge 
 which has lead to
 important developments and understanding of this theory as related to strings
 \cite{polya}.Surprisingly enough, though gravity is  more complicated than 
 Yang-Mills gauge theory, it turned out that the $4$D gravitational effective 
 action could have been derived starting from the trace anomaly itself 
 \cite{rig}. 
 On the other hand, to our best knowledge, apart from  
 mentioned $2$D examples, the form of the $4$D or higher dimensional anomalous
 gauge effective action is not present in the literature.  Fortunately, 
 difficulty related to the calculation of the effective action have not 
 hindered study of anomalies since methods of anomaly calculation have 
 been designed in such a way to avoid explicit knowledge of the effective 
 action, so that divergences of the  anomalous currents,  were 
 extracted from {\it  local counter-terms,} \cite{jackiw}, \cite{fujikawa},
 \cite{hkern}.\\

 In this paper, we would like to present 
 the complete, anomalous, effective action for an Abelian
 gauge field theory coupled  to Dirac fermions in even dimensions
 $D=2k$.\\
 To properly pave our way let us briefly recall the origin of gauge anomalies.
  Dirac fermions coupled  to external gauge fields can be integrated out in
  a path integral and give the anomalous gauge field
  effective action induced by  quantum effects. We  choose to depart
  from a standard treatment of axial anomaly where chiral symmetry is a {\it 
  global invariance } of the theory. Namely,  fermions will posses {\it
  local} axial symmetry, so we introduce both vector and axial vector gauge
  fields in the covariant derivative. This approach is crucial in order to be
  able to construct effective action from which axial and vector current
  can be recovered.  \\
  Guidelines for construction of anomalous effective action are potentially
  anomalous Feynman diagrams, which in $2k$ dimensions, have $k+1$ fermion
  propagators \cite{frampt} with $2m +1 $ axial vertices and $k-2m$ vector
  gauge field vertices. The odd number of  vertices with axial vector gauge
  field follow from the condition $\left(\gamma^5\right)^{2m+1}=\gamma^5$ and
  therefore such diagrams have the same internal structure as those with one 
  axial vertex.  On the
  basis of the above reasoning we found  the general form of the anomalous 
  gauge effective action in $D=2k$ dimensions to be

\begin{eqnarray}
\Gamma=\sum_{m=0}^{m_{max}}{ g^{2m+1} e^{k-2m}\over (2\pi)^k }\int d^{2k}x\,  
\epsilon^{\mu_1\dots \mu_{2k-4m}\nu_1\dots\nu_{4m} }
F_{\mu_3\mu_4}\dots\times
F_{\mu_{2k-4m-1}\mu_{2k-4m}}&&\times\nonumber\\
F_{\nu_1\nu_2}^5\times\dots\times F_{\nu_{4m-1}\nu_{4m}}^5
\left[\, F_{\mu_1\mu_2}{1\over\Box }
\left(\,\partial^\mu A^{5}_\mu\,\right)  
- 2{\bf a}\,  A_{\mu_1}A^5_{\mu_2}\,\right]&&\label{effact}
\end{eqnarray}

Short explanation is needed for the meaning of $m_{max}$. The value
of the integer $m$ is restricted by $m\le (k-1)/2$,  thus $m_{max}$
represents the maximal {\it integer }  obtained from such relation.\\
From general arguments, as well as from explicit calculations, one expects 
that the extraction of anomalies is subject to regularization arbitrariness.
This regularization dependence is seen in (\ref{effact})  through the presence
of an arbitrary parameter ${\bf a}$  in the {\it local}
part of the effective action \cite{noi}. The axial gauge current obtained
from (\ref{effact}) is

\begin{eqnarray}
&&J^{5\, \mu}\equiv {1\over g}{\delta S\over\delta A^{5}_\mu}   \nonumber\\
&& = -\sum_{m=0}^{m_{max}}{ g^{2m} e^{k-2m}\over (2\pi)^k }
 \left[\, 
\epsilon^{\mu_1\dots \mu_{2k-4m}\nu_1\dots\nu_{4m}}
{\partial^\mu\over\Box}\left(\,
F_{\mu_1\mu_2}\times\dots\times
F_{\mu_{2k-4m-1}\mu_{2k-4m}}F_{\nu_1\nu_2}^5\times\dots\times 
F_{\nu_{4m-1}\nu_{4m}}^5
\,\right) 
\right.
\nonumber\\
&& -4m
\left. 
\epsilon^{\mu\mu_2\dots \mu_{2k-4m+2}\nu_3\nu_4\dots\nu_{4m}}
F_{\mu_3\mu_4}\times\dots\times F_{\mu_{2k-4m+1}\mu_{2k-4m+2}}
F_{\nu_3\nu_4}^5\times\dots\times F_{\nu_{4m-1}\nu_{4m}}^5
{\partial_{\mu_2}\over\Box}\partial A^5
\right.
\nonumber\\
&& 
\left.
+\epsilon^{\mu\mu_2\dots \mu_{2k-4m}\nu_1\dots\nu_{4m}} 
 F_{\mu_3\mu_4}\times\dots\times
F_{\mu_{2k-4m-1}\mu_{2k-4m}}                                  
F_{\nu_3\nu_4}^5\times\dots\times F_{\nu_{4m-1}\nu_{4m}}^5\times
\right.
\nonumber\\
&&
\left.
\times\left(\, 8\,{\bf a}\, m \,F_{\nu_1\nu_2}\, A_{\mu_2}^5 + 2\,{\bf a}\,
m(1-4m)\,F_{\nu_1\nu_2}^5\, A_{\mu_2}\,\right)\,
\right]\label{axial} 
\end{eqnarray}

from (\ref{axial}) one finds  anomalous divergence of the axial currents as

\begin{eqnarray}
\partial_\mu J^{5\, \mu}=({\bf a}\,-1)\,\sum_{m=0}^{m_{max}}{ g^{2m} e^{k-2m}
\over (2\pi)^k }&& \epsilon^{\mu_1\dots \mu_{2k-4m}\nu_1\dots\nu_{4m}}
\, F_{\mu_1\mu_2}\times\dots\times \nonumber\\
&& F_{\mu_{2k-4m-1}\mu_{2k-4m}}
F_{\nu_1\nu_2}^5\times\dots\times F_{\nu_{4m-1}\nu_{4m}}^5
\end{eqnarray}

On the other hand, the vector current obtained  from (\ref{effact}) is

\begin{eqnarray}
J^{\mu}\equiv && {1\over e}{\delta S\over \delta A_\mu} 
= \sum_{m=0}^{m_{max}}{ g^{2m+1} e^{k-2m-1}\over (2\pi)^k }
\epsilon^{\mu_1\dots \mu_{2k-4m}\nu_1\dots\nu_{4m}}
\, F_{\mu_5\mu_6}\times\dots\times F_{\mu_{2k-4m-1}\mu_{2k-4m}}
\nonumber\\
\times && F_{\nu_1\nu_2}^5\times\dots\times F_{\nu_{4m-1}\nu_{4m}}^5
\left[\,
(k-2m)\, F_{\mu_3\mu_4}{\partial_{\mu_2}\over\Box}\left(\,
 \partial^\rho A^{5}_\rho\, \right)
 \right.
 \nonumber\\ 
  - && 
  \left.
 {\bf a}\,(k-2m)\,  F_{\mu_3\mu_4}
A^{5}_{\mu_2}+ {\bf a}\,(k-2m-1)F_{\mu_3\mu_4}^{5} A_{\mu_2} \,
\right]
\end{eqnarray}

which leads to the vector gauge current anomalous divergence as 

\begin{eqnarray}
\partial_\mu J^{ \mu}= &&-{\bf a}\sum_{m=0}^{m_{max}}{ g^{2m+1} e^{k-2m-1}
\over (2\pi)^k}
 \epsilon^{\mu_1\dots \mu_{2k-4m}\nu_1\dots\nu_{4m}}
\,F_{\mu_1\mu_2}^5 F_{\mu_3\mu_4}\times\dots\times  \nonumber\\
\times && F_{\mu_{2k-4m-1}\mu_{2k-4m}}
F_{\nu_1\nu_2}^5\times\dots\times F_{\nu_{4m-1}\nu_{4m}}^5 
\end{eqnarray}

In the above result one recognizes a well known pattern of incompatibility of
classical symmetries, i.e. no choice of the parameter ${ \bf a}$ can preserve 
both symmetries at the quantum level. To be truthful, it is the 
divergence of the axial current which  normally appears in the
literature \cite{jackiw},\cite{frampt}. From the above result it is clear why, 
i.e. all explicit method of calculation { \it assume} gauge invariant 
regularization which in our
language implies the choice ${\bf a}=0$. Furthermore, previous considerations
were limited to global axial symmetry which amounts to setting all axial
gauge fields couplings to zero. We strongly suggest the use of the 
parameter ${\bf a}$ in the formulae, and only { \it a posteriori} choose a 
particular value, i.e. a definite regularization scheme. In this way one can 
easily track the interplay and incompatibility of the classical symmetries
at the quantum level. Furthermore, it permits easy shift from one local symmetry
to another, no matter how complicated the  expressions for currents are.\\

In order to make the general formalism more transparent, as well as to
verify the agreement with known results, let us start from $2$D. From 
(\ref{effact}), choosing $k=1$, one finds effective action as

\begin{equation}
S={ e g\over 2\pi }\int d^{2}x\,  \epsilon^{\mu\nu }\,
\left[\, F_{\mu\nu}{1\over\Box }\left(\,\partial^\rho A^{5}_\rho\,\right)
-2{\bf a}\, A_{\mu}A^{5}_{\nu}\,\right]
\label{qed2}
\end{equation}

Exploiting a  $2$D {\it identity} $A_\mu = \epsilon_{\mu\nu}A^{5\,\nu}$ ($e=g)$, 
transforms (\ref{qed2}) in the well known form of the Schwinger effective 
action

\begin{equation}
S=-{ e^2\over 4\pi }\int d^{2}x\, \left[\, F_{\mu\nu}{1\over\Box }\, F^{\mu\nu}
+2{\bf a}\, A^{\mu}A_{\mu}\,\right]
\end{equation}

together with the anomalous divergences of vector and axial vector current 

\begin{eqnarray}
&&\partial_\mu J^\mu= -{\bf a}\,{e\over\pi}\,\epsilon^{\mu\nu }\, \partial_\mu 
A_\nu^{5}
                    = -{\bf a}\,{e\over\pi}\, \partial_\mu A^\mu\\
&&\partial_\mu J^{5\, \mu}=  ({\bf a}-1)\, {e\over\pi }\, \epsilon^{\mu\nu }\, 
\partial_\mu A_\nu
\end{eqnarray}              

To proceed a  step further, we take $k=2$ in (\ref{effact}) and  find, to our 
best knowledge, so far unreported $4$D effective  anomalous action 

\begin{equation}
S={ g e^2 \over 4\pi^2 }\int d^4x\,  \epsilon^{\mu\nu\alpha\beta }\,
\left[\, F_{\mu\nu}\, F_{\alpha\beta } \,  {1\over\Box }\,
\left(\,\partial^\rho A^{5}_\rho\,\right)
- 2{\bf a}\,A_{\mu}\, A^{5}_{\nu}\, F_{\alpha\beta }\,\right]
\end{equation}

which enables us to find { \it explicit } form of the vector and axial vector 
currents as

\begin{eqnarray}
&& J^{\mu}={ eg\over 4\pi^2 }\,\epsilon^{\mu\nu\alpha\beta }\, \left[\,
\, F_{\alpha\beta}\, {\partial_{\nu}\over\Box}\left(\,
 \partial^\rho A^{5}_\rho\, \right) - 2{\bf a}\, F_{\alpha\beta}\,
A^{5}_{\nu}+{\bf a}\,\, F_{\alpha\beta}^{5}\, A_{\nu}\,\right]\label{j}\\
&& J^{5\, \mu}=-{ e^2\over 4\pi^2 }\,\epsilon^{\rho\nu\alpha\beta }\,\left[\,
{\partial^{\mu}\over\Box}\left(\,F_{\rho\nu}\, F_{\alpha\beta}\right)+2{\bf a}\,
\delta^\rho_\mu\,  F_{\alpha\beta}\, A_\nu\,\right]\label{j5}
\end{eqnarray}

In spite of the complicated form of currents (\ref{j}) and  (\ref{j5}), one 
finds simple looking, well known, $4$D divergences 

\begin{eqnarray}
&&\partial_\mu J^{\mu}= -eg\,{{\bf a}\,\over 
4\pi^2}\,\epsilon^{\mu\nu\alpha\beta }
\,F_{\mu\nu}\, F_{\alpha\beta}^5\label{divj} \\
&&\partial_\mu J^{5\mu}= e^2\,{({\bf a}\,-1)\over 4\pi^2}\,
\epsilon^{\mu\nu\alpha\beta }\,
F_{\mu\nu}\, F_{\alpha\beta}\label{divj5}
\end{eqnarray}

We shall also exhibit  a $6$D ($k=3$) results because it is the first dimension
in which more than one axial vertex can be present. In fact, one finds
$m=0$ (one axial vertex),  $m=1$ (three axial vertices). They give the
anomalous action as

\begin{equation}
\Gamma={ g e\over (2\pi)^3 }\int d^{6}x\,  
\epsilon^{\mu\nu\rho\sigma\tau\omega }\left(\, 
e^2 F_{\rho\sigma}F_{\tau\omega } + g^2F_{\rho\sigma}^5F_{\tau\omega
}^5\,\right)
\left[\, F_{\mu\nu}{1\over\Box } \partial A^{5} 
- 2{\bf a}\,  A_{\mu}A^5_{\nu}\,\right]
\label{effact6}
\end{equation}

(\ref{effact6}) leads to the anomalous divergences of currents as

\begin{equation}
\partial_\mu J^{ \mu}= -g{{\bf a}\over
(2\pi)^3}\epsilon^{\mu\nu\rho\sigma\tau\omega }F_{\mu\nu}^5\left[\,
  e^2\, F_{ \rho\sigma}  F_{\tau\omega }
 + g^2  F_{ \rho\sigma}^5  F_{\tau\omega }^5 \right]
\label{divj6}
\end{equation}

\begin{equation}
\partial_\mu J^{5 \mu}= e{{\bf a}-1\over
(2\pi)^3}\epsilon^{\mu\nu\rho\sigma\tau\omega }F_{\mu\nu}\left[\,
  e^2\, F_{ \rho\sigma}  F_{\tau\omega }
  + g^2  F_{ \rho\sigma}^5  F_{\tau\omega }^5 \right]
\label{divj56}
\end{equation}

In (\ref{divj6}) and (\ref{divj56}) we see the appearance of  axial
vertices, in particular the difference 
in the axial current conservation compared to \cite{frampt}. The presence
of additional $ F_{ \rho\sigma}^5$  is caused by  the local axial symmetry 
considered in this paper.\\

In summary, we  have constructed anomalous effective
action for Abelian gauge fields which represents higher dimensional extension
of $2$D Schwinger model. To our satisfaction, the anomalous effective action  
turns out to have  particularly simple form whose kernel is a $2$D effective 
action written in  terms  of {\it independent} vector and axial vector gauge 
fields.The higher dimensional input is {\it all} encoded in the sequence of 
vector and axial vector gauge field strengths which contract remaining
indices in the $2k$-dimensional Levi-Civita tensor.

\end{document}